\documentclass[twocolumn,prl,floatfix,showpacs]{revtex4}
\usepackage{color}
\usepackage{epsfig}
\usepackage{graphicx}
\usepackage{amsmath}

\begin{document}

\title{ Velocity selection problem in the presence of the triple junction}

\author{E. A. Brener}
\author{C. H\"uter}
\author{D. Pilipenko}
\author{D. E. Temkin}
\affiliation{Institut f\"ur Festk\"orperforschung, 
Forschungszentrum J\"ulich, D-52425 J\"ulich, Germany}


\date{\today}

\begin{abstract}
Melting of a bicrystal along the grain boundary is discussed.  
A triple junction plays a crucial role in the velocity selection problem in this case. 
In some range of the parameters an entirely analytical solution of this problem is given.
 This allows to present a transparent picture of the structure of the selection theory. 
We also discuss the selection problem in the case of the growth 
of a ``eutectoid dendrite'' where a triple junction is present 
because three phases are involved in the eutectoid reaction.

\end{abstract}

\maketitle

During the last decades, our understanding of pattern formation in various nonlinear 
dissipative systems has made a remarkable progress. Building on these foundations,  
it has now become possible to develop a description of a large class of patterns that 
are found in diffusional growth. 
In the two-dimensional dendritic growth the 
needle crystal is assumed to be close to the parabolic Ivantsov solution \cite{Ivantsov}.
If anisotropic capillary effects are included, then a single dynamically stable solution 
is found for any external growth condition.  
In this classical problem of dendritic growth, velocity selection is controlled 
by tiny singular effects of the anisotropy of the surface energy
(see, for example, \cite{kessler}, \cite{brenermelnikov}). In the case of isotropic 
surface energy, the dendritic solution does not exist and instead the so-called doublon 
structure is the solution of the problem \cite{benamar95,ihle93}.
The dendritic-like structures which occur during melting have also been  observed experimentally 
(see \cite{gor93} and references therein). 

Moreover, for many applications, such as grow of dislocation-free silicon crystals \cite{wang}, the investigation of the influence of the defects on the melt-crystal interface behavior  
is of great importance. 
The melting along  defects such as grain boundaries in inhomogeneous materials is typically favorable, since
they serve as predecessor to melting. 
In this paper we therefore discuss the melting process along the grain boundary in the crystal
(see Fig. 1). 
The presence of a triple junction at the tip of the melting zone  
leads to an entirely different selection mechanism, since the triple junction produces 
a very strong perturbation of the liquid-solid interfaces and weak anisotropy effects 
can be neglected. This idea was expressed in \cite{tem}. 
We also discuss importance of a triple junction for the selection problem 
of the growth of a ``eutectoid dendrite''  (see Fig. 2).  
\begin{figure}
\begin{center}
\epsfig{file=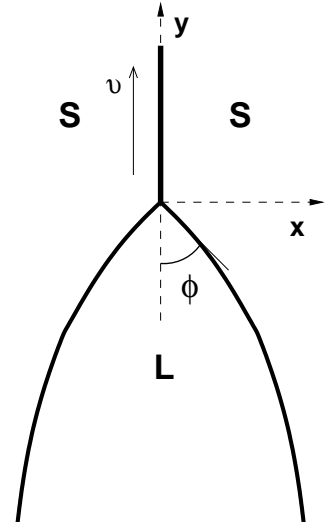, angle=0, height=7cm} 
\caption{Propagation of the melt zone along the grain boundary ($x=0$). The melting 
front propagates with velocity $\upsilon$ along the y-direction. S corresponds to 
two solid grains and  L corresponds to the liquid phase. 
The triple junction is at the origin of the coordinate system. 
} 
\label{force_eq}
\end{center}
\end{figure}

\begin{figure}
\begin{center}
\epsfig{file=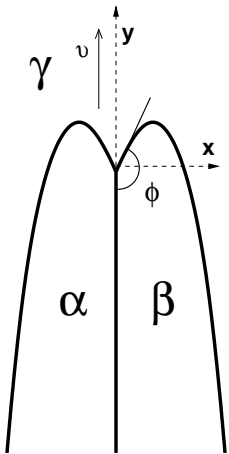, angle=0, width=3cm} \caption{Schematic picture of the interface 
structure  in the eutectoid reaction.
The structure  propagates with a constant velocity 
$\upsilon$ in the $y$ direction. The triple junction is located at the origin of 
the coordinate system. 
} \label{eutectic}
\end{center}
\end{figure}

{ \it Melting along a grain boundary.}
We consider the two dimensional melting problem of a pure bicrystal with a straight 
grain boundary as shown in Fig. \ref{force_eq}. 
Overheating of the crystal leads to a moving melt zone. 
The contact angle $\phi$ at the triple junction is related to the surface energies 
by Young's law: $2 \gamma \cos \phi = \gamma_b$, where $\gamma$ is the surface energy 
on the liquid-solid interface  and $\gamma_b$ is the surface energy of 
the grain boundary.
We introduce the dimensionless temperature field $u=c_p(T_{\infty}-T)/L$, where 
$L$ is the latent heat,  $c_p$ is heat capacity, and $T_\infty$ is the temperature 
in the crystal far away from the interface.  

The temperature field $u$ obeys the following heat diffusion equation and 
boundary conditions:
\begin{equation}
\label{heatdiff}
D\nabla^2 u=\partial u/\partial t,
\end{equation}
\begin{equation}
\label{energy}
\upsilon_n= D\vec{n}(\vec{\nabla} u_L |_{int}-\vec{\nabla} u_S |_{int}),
\end{equation}
\begin{equation}
\label{localeq}
 \Delta-d_0\kappa=u|_{int}, 
\end{equation}
where L and S refer to the liquid and solid, respectively,   
$d_0 = \gamma T_M c_p/L^2$ is the capillarity length,  
$T_M$ is the melting temperature, and 
$\kappa$ is the curvature of the liquid-solid interface, which assumed to be 
positive for convex interfaces (as in Fig. 1).
 We  introduce the dimensionless overheating 
$\Delta = c_p(T_{\infty}-T_M)/L$, and the Peclet number 
$p =  \upsilon R/2 D$ with $D$ being the thermal diffusion constant 
(assumed to be the same in both phases), 
$R$ is the radius of curvature of asymptotically 
fitted Ivantsov parabola and $\upsilon$ is
the velocity of the steady-state motion.

The physics underlying  Eqs.~(\ref{heatdiff}-\ref{localeq}) is quite simple.
A moving melting front absorbs the latent heat. Requirement of the heat conservation 
at the interface gives us Eq.~(\ref{energy}) (${\bf n}$ is the normal to the front 
pointing into the solid phase; $\upsilon_n$ is the normal velocity of the front). 
The local thermodynamical equilibrium in the interfacial region implies 
Eq.~(\ref{localeq}). This is the Gibbs-Tomson relation, which gives the equilibrium 
value of the temperature at the interface taking into account curvature corrections. 

An equivalent formulation of the problem which is more convenient for our purposes is 
obtained by eliminating the thermal field. This can be done by using the standard 
Green's function techniques to obtain an  equation in closed form for the shape 
of the solid-liquid interface:  
 
\begin{equation}
\label{full_nonlin_main}
\Delta(p) - \frac{d_0}{R}\kappa = \frac{p}{\pi} \int_{-\infty}^{\infty}dx' e^{-p(y - y')} K_0 (p\eta(x, x')),
\end{equation}
where $\eta = [(x-x')^2+(y-y')^2]^{\frac{1}{2}}$, and  $K_0$ is the modified Bessel function of third kind in zeroth order; all lengths are measured in units of $R$. 
The relation between the  overheating $\Delta$ and the Peclet number $p$ is given 
by the Ivantsov relation: $\Delta = \sqrt{p \pi} \exp{(p)} \text{ erfc}(\sqrt{p})$.

Eq. (\ref{full_nonlin_main}) is a complicated nonlinear integro-differential equation. 
We should find a smooth solution of this equation which has a proper angle at the tip 
and which is close to the Ivantsov parabola ($y=-x^2/2$) in the tail region. 
The classical dendritic growth problem  does not have a solution with isotropic 
surface tension and with  a smooth tip which corresponds to the angle $\phi=\pi/2$ 
\cite{kessler,brenermelnikov}. This statement can be expressed in the following form. 
For any given positive values of the Peclet number $p$ 
and the so-called stability parameter 
$\sigma=d_0/pR$ the symmetric solution which is close to the Ivantsov parabola in the 
tail region has an angle at the tip $\phi=f(\sigma,p)< \pi/2$. The limit $\sigma=0$ and 
$\phi=\pi/2$ is a singular limit for that problem.  
For example, Meiron \cite{meiron} numerically calculated the angle $\phi$ as a function
of $\sigma$ for several values of the Peclet number with isotropic surface tension 
and found that the angle $\phi$ remains smaller than $\pi/2$ for any positive $\sigma$. 
  
Our problem  differs from the classical dendritic problem precisely in  that we need to 
satisfy the condition that the angle $\phi$ at the tip should be smaller than $\pi/2$. 
This means that we can select the stability 
parameter $\sigma=\sigma^{\star}(\phi,p)$ as a function of $\phi$ and $p$.   
In order to illuminate this selection mechanism  we consider the case of small 
opening angles i.e., $\phi\ll 1$ and small overheating ($p\ll1$) which allows us to 
drastically simplify the problem and 
finally obtain the selection in some limit entirely analytically.

\begin{figure}
\begin{center}
\epsfig{file=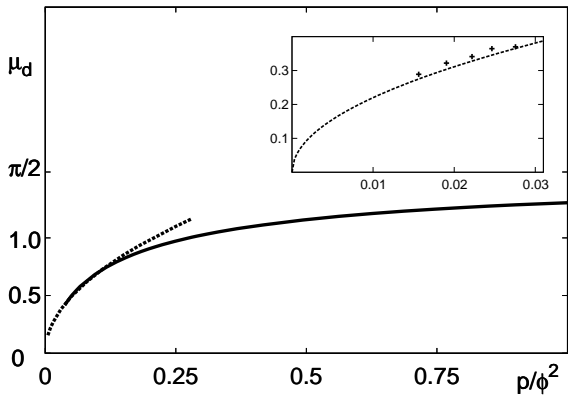, angle=-90, width=8cm}
\caption{Eigenvalue $\mu_d$ as a function of  $p/\phi^2$. Solid line represents the 
numerical solution of Eq.~(\ref{approx_lin_expl}), dash line corresponds to the 
asymptotic behavior $\mu_d= 2.2\sqrt{p/\phi^2}$.
The inset shows an enlarged region of small values of $p/\phi^2$, the dashed line is $2.2\sqrt{p/\phi^2}$, crosses correspond to the data extracted from  Ref.\cite{meiron}.}
\label{LinGamma}
\end{center}
\end{figure}

For small opening angles $\phi$ the following rescaling of coordinates 
is useful in order 
to eliminate small parameters from the interface shape: 
\begin{eqnarray}
x\rightarrow x/\phi, \quad y\rightarrow y/\phi^2.
\end{eqnarray}
After this rescaling, $y=-|x|$ near the tip and  in the tail region 
the asymptotic of the Ivantsov parabola ($y=-x^2/2$) remains unchanged.  
 In this small angle approximation the function  $\eta$ depends only on $y$ 
variable,  $\eta\approx |y-y'|\phi^{-2}$, because $x\ll y$.
 Consequently the whole integral kernel in  
Eq.~(\ref{full_nonlin_main}), under the assumption of small Peclet numbers, is function  
only of $y$.  
 Due to the symmetry of the interface we consider it only for $x>0$. 
It is more convenient to treat the shape as a function $x(y)$ and to make change of variables, 
 $$y\rightarrow -y.$$
 The curvature of the interface can be written as $\kappa\approx- d^2x/dy^2$.   
Performing all these step we reduce the original {\it nonlinear}  
 Eq.~(\ref{full_nonlin_main}) to the following {\it linear} equation:
\begin{equation}
\label{approx_lin_expl}
1+\mu_d \frac{d^2 x}{dy^2} = \frac{2}{\pi} \sqrt{\frac{ p}{\pi\phi^2}} \int_0^{\infty} dy' \frac{dx(y')}{dy'} e^{-\frac{p}{\phi^2}(y'-y)} K_0(\frac{p}{\phi^2}|y'-y|),
\end{equation}
where $\mu_d=d_0\phi^3/R\Delta =\sigma\phi^3\sqrt{p/\pi} $. 
As  follows from Eq.~(\ref{approx_lin_expl}), the eigenvalue $\mu_d$ is a function 
 only of one parameter, $p/\phi^2$.
We solve Eq. (\ref{approx_lin_expl}) numerically using  standart linear algebra routines. The results are presented in Fig. \ref{LinGamma}.

For small values of $p/\phi^2 \ll 1$,  the eigenvalue $\mu_d$ is proportional to the 
 prefactor in front of the integral term in Eq.~(\ref{approx_lin_expl}) i.e., 
$\mu_d = \alpha \sqrt{(p/\phi^2)}$. 
We found numerically that $\alpha\approx 2.2$, which leads to the following 
scaling relations   for the selected tip radius and velocity: 
\begin{equation} \label{MainRadLin}
\frac{R}{d_0} \approx 0.81 \phi^4 \Delta^{-2}, \quad \frac{\upsilon d_0}{D} \approx 0.79 \phi^{-4} \Delta^4.
\end{equation}
We compare our asymptotic predictions $\mu_d=2.2\sqrt{p/\phi^2}$, with data extracted  from Fig. 1c of the Ref.~\cite{meiron}. Despite the fact that the results of Ref.~\cite{meiron} were obtained by the solution of the full Eq.~\ref{full_nonlin_main}, and our results were obtained under the assumption of  small opening angles, they show a relatively good quantitative agreement up to angles $\phi\approx \pi/6$ (see the inset in  Fig.~\ref{LinGamma}).

In the case of large value of $p/\phi^2\gg 1$ further simplifications of Eq.~(\ref{approx_lin_expl}) are  possible.
 The combination of the exponential term with the asymptotic of the Bessel 
function for the large arguments leads  to the integral kernel 
$1/\sqrt{p\phi^{-2}(y-y')}$ for the argument $y'<y$, 
 while, for  $y>y'$ the integral kernel is exponentially small and can be neglected. 
Thus,  Eq.~(\ref{approx_lin_expl}) finally reads 
\begin{equation}
\label{limitcase_eq}
1 + \mu_d \frac{d^2x}{dy^2}=
 {\frac{\sqrt{2}}{\pi} \int_0^y
 \frac{dx(y')}{dy'}dy' \frac{1}{\sqrt{ y-y'}}}.
\end{equation}
 This equation for the unknown function $f(y)=dx/dy$ can be solved by means of 
the Laplace-transformation 
\begin{eqnarray*}
 F(s)=\int_0^{\infty}f(y)\exp(-sy)dy.
\end{eqnarray*}
Using the boundary condition, $f(0)=1$, we get
\begin{equation}
\label{LaPlaceTransform}
F(s) = \frac{\mu_d \sqrt{s}-1/\sqrt{s}}{\mu_ds^{3/2} - \sqrt{2/\pi}}.
\end{equation}
The function F(s) has a pole at $s = (2/\pi \mu_d^2)^{1/3}$. Therefore the 
physically relevant solution exists only if
the parameter  $\mu_d = \pi/2$, which means that the pole and zero of the  
function $F(s)$ coincide.  
To our best knowledge this is the first example where the problem of 
 velocity selection in dendritic growth is 
solved entirely analytically. This exact analytical treatment demonstrates in a 
very transparent way why the physically relevant solution 
of Eq.(\ref{LaPlaceTransform}) exists only for some specific value  
of the  parameter $\mu_d$. 
The inverse Laplace-transform of Eq.(\ref{LaPlaceTransform}) finally gives 
the closed expression for  the derivative of the interface profile:
\begin{align*}
 &\frac{dx}{dy} =\left(\frac{1}{2} + \frac{i}{2\sqrt{3}}\right)\exp\left(\frac{(-1+i\sqrt{3})y}{\pi}\right) \times \\  &\times \text{erfc}\left[\left(\frac{1+i\sqrt{3}}{\sqrt{2\pi}}\right)\sqrt{y}\right] + c.c.
\end{align*}

We note that the same analytical treatment is possible also in the case where 
interface kinetics effects are included. The interface kinetics requires an 
additional term, $-\upsilon_n/\beta$, in the left-hand-side of 
Eq. (\ref{full_nonlin_main}) and, correspondingly, the term $-\mu_{\beta}dx/dy$
in Eqs. (\ref{approx_lin_expl}, \ref{limitcase_eq}) ($\beta$ is the kinetic coefficient
 and $\mu_{\beta}=\upsilon\phi/(\beta\Delta)$). In this case the selection condition 
reads $\mu_d=\pi(1-\mu_{\beta})^2/2$.   

{\it Eutectoid "dendrite``}

\begin{figure}
\begin{center}
\epsfig{file=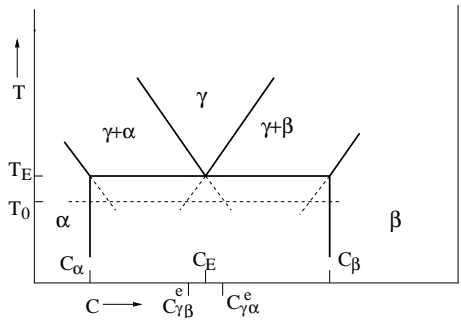, angle=0, width=6cm} \caption{ The schematic phase diagram of 
the eutectoid reaction.
$T_0$ is the temperature of the system, $T_E$ is the eutectoid temperature. 
The regions $\gamma$, $\alpha$ and $\beta$ correspond to the one phase equilibrium 
state. $\gamma+\alpha$ and $\gamma+\beta$ are regions of two-phase equilibrium. 
$C_{\alpha}$, $C_{\beta}$ and $C_E$ are the equilibrium concentrations at the eutectoid
temperature. For this diagram the eutectoid composition $C_E=(C_{\alpha}+C_{\beta})/2$; 
the lines of the $\gamma-\alpha$ phase equilibrium are parallel, the same for 
lines of $\gamma-\beta$ equilibrium; the equilibrium concentrations at the $\alpha-\beta$ 
interface, $C_{\alpha}$ and $C_{\beta}$ do not depend on the temperature.}   
\label{phase}
\end{center}
\end{figure}

Another illustrative example of selection by the triple junction is the eutectoid reaction. 
We consider here a rather simple model of the eutectoid system, described by the phase diagram in Fig.~\ref{phase},  in order to obtain the equation in the spirit of Eq.~(\ref{full_nonlin_main}). 
\begin{figure}
\begin{center}
\epsfig{file=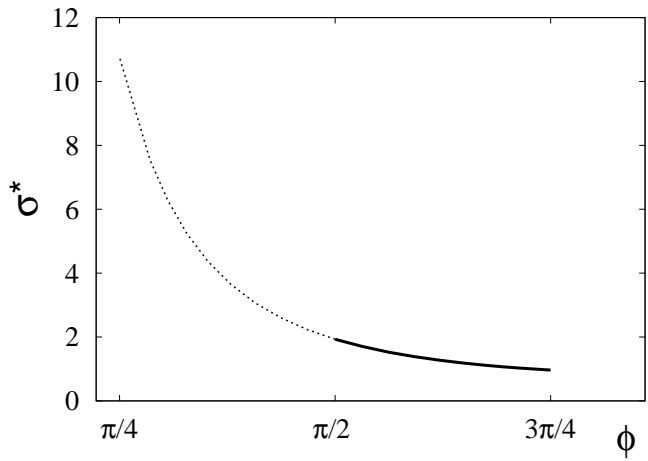, angle=-90, width=9cm} \caption{ Eigenvalue $\sigma$ as a 
function of the opening angle.
 The solid line ($\phi>\pi/2$) corresponds to the growth of $\alpha$ and $\beta$ phases
from the $\gamma$ phase, $T_0<T_E$.
 The dotted line ($\phi<\pi/2$) to the inverse process, $T_0>T_E$.
 } \label{angle}
\end{center}
\end{figure} We restrict the consideration to the case of an isolated eutectoid 
dendrite (Fig. 2), i.e., we consider the situation when the period of the 
lamellar structure is much larger than the characteristic tip scale of 
the "dendrite``. 
We call the appearing structure, shown in Fig.~\ref{eutectic}, "dendrite" because, 
as we show later, it has the parabolic asymptotics. 
The concentration fields have jumps on the phase boundaries in eutectoid reactions. 
As a consequence the difference of concentration $C_E-C_{\alpha}$ (source term), 
and  $C_E-C_{\beta}= (C_E-C_{\alpha})$ (sink term) appear as the amplitudes 
in front of the integral terms. 

Moreover,  we also assume that the diffusion constants are equal in all three solid phases. 
This assumption crucially simplifies the model.
The concentration $C_{\gamma}$ far away from the interface is chosen to be $C_E$ leading to the symmetry of the shape as presented in Fig. 2.
Finally, this system can be described by the integro-differential equation
 as in the case of classical dendritic growth. Due to the assumed symmetry of the 
phase diagram 
and of the shape of the interfaces, we  consider  only positive values of $x$, i.e., 
write down the equation only for the $\gamma-\beta$ interface:  
\begin{eqnarray}
\Delta - \frac{d_0}{R}\kappa =-\frac{p}{\pi} \int_{-\infty}^{0} dx' e^{-p(y - y')} 
K_0 (p\eta (x, x'))\nonumber\\
+\frac{p}{\pi} \int_{0}^{\infty} dx' e^{-p(y - y')} K_0 (p\eta(x, x')).
\label{eutect_eq}
\end{eqnarray}
Here $d_0$ is the capillary length \cite{kassner}, $\kappa$ is the curvature of the 
interface which is positive in Fig. 2,  and 
$\Delta=(C_E-C_{\gamma\beta}^{e})/(C_{\beta}-C_E)$ is the supersaturation. 
This equation  differs from Eq.~(\ref{full_nonlin_main}) only by the sign in front 
of the first integral, which represents the source at the $\alpha-\gamma$ interface, 
while the second integral corresponds to the sink at the $\beta-\gamma$ boundary. 
In other words, this simple modification of the classical equation of dendritic growth  
already contains the key ingredient of  the eutectoid reaction.

In the most relevant case of small Peclet numbers this equation  can be further 
simplified.
We stress that the  usual relation between the Peclet number and the 
supersaturation $\Delta \approx\sqrt{\pi p}$ ,  
is no longer valid here. 
In the small Peclet number  limit the flux in the  $\alpha$ and $\beta$ phases 
is much larger than in the $\gamma$ phase. Thus, in the asymptotics, one can write 
 $D\Delta/x=-\upsilon dx/dy$. This leads to the parabolic asymptotics of the shape, 
$y=-x^2/2R$, and to the  
following relation between the  supersaturation and the Peclet number, 
$p=\upsilon R/2D$:
$$\Delta=2p.$$
Thus, for small supersaturation $\Delta$  Eq.~(\ref{eutect_eq}) reads:
\begin{eqnarray*}
2 - \sigma \kappa = \frac{1}{2\pi} \int_{0}^{\infty} dx' \ln \left
( \frac{(x+x')^2+(y-y')^2}{(x-x')^2+(y-y')^2}\right),
\end{eqnarray*}
where $\sigma=d_0/Rp$.
Our purpose is to find the shape with parabolic asymptotics and  the eigenvalue $\sigma$ 
for the given angle $\phi$. 
We solve this nonlinear equation  numerically in the spirit of Ref~\cite{meiron}. 
The resulting eigenvalue $\sigma=\sigma^{\ast}(\phi)$ as  function of the opening angle 
is shown in  Fig.~\ref{angle}.
We notice,  that in contrast to the problem of classical dendritic growth, 
where $\sigma=\sigma^{\ast}(\phi,p)$ has a singular point at $\phi=\pi/2$, 
in the case of the eutectoid reaction, a solution exists even for  $\phi=\pi/2$.
Angles $\phi>\pi/2$ correspond to growth of  $\alpha$ and $\beta$ phases from 
the $\gamma$ phase ($T_0<T_E$), while angles $\phi<\pi/2$ describe the inverse case 
($T_0>T_E$). 
In the latter case the structure of the interfaces is the same as presented in 
Fig.~\ref{force_eq}, with the  replacement of  $L\rightarrow \gamma$ and 
$S\rightarrow \alpha,\beta$. The related model of  melting of  eutectic structures was 
discussed in \cite{brener2007}.

Since the eigenvalue $\sigma^{\ast}$ is found, the velocity of the process can be 
written as $\upsilon=2 \sigma^{\ast} D p^2/d_0$. 
Consequently, the velocity of eutectoid growth is proportional to $\Delta^2$, 
while for the classical dendritic growth it scales as  $\Delta^4$.

In conclusion, the selection mechanism due to the presence of the triple junction is very different 
from the anisotropy effects which are responsible for the selection in the classical 
theory of dendritic growth. 
In the  regime of small opening angles the problem  
of  velocity selection in melting along the grain boundary has been 
solved entirely analytically. This exact analytical treatment represents in a   
very transparent way the structure of the selection theory.
We hope that results of this paper will stimulate some new experiments 
to bring new insights to this interesting problem. 

We acknowledge the support by the Deutsche Forschungsgemeinschaft under the Project 
MU 1170/4-1.

\end{document}